\documentclass{epl}
\usepackage{epsf}

\begin{document}

\title{Strong Evidence of Normal Heat Conduction \protect \\
in a one-Dimensional Quantum System}
\shorttitle{Strong Evidence of Normal Heat Conduction}
\author{Keiji Saito}
\shortauthor{K. Saito}
\institute{Department of Applied Physics, School of Engineering \\ 
University of Tokyo, Bunkyo-ku, Tokyo 113-8656, Japan}


\maketitle
\begin{abstract}
We investigate how the normal energy transport is realized 
in one-dimensional quantum systems using a quantum spin system.
The direct investigation of local energy distribution 
under thermal gradient is made using the quantum master
equation, and the mixing properties and the convergence of 
the Green-Kubo formula are investigated when the number of spin increases. 
We find that the autocorrelation function in the Green-Kubo formula
decays as $\sim t^{-1.5}$ to a finite value which
vanishes rapidly with the increase of the system size. 
As a result, the Green-Kubo formula converges to a finite value 
in the thermodynamic limit. These facts 
strongly support the realization of Fourier heat law in a quantum system.
\end{abstract}

The Fourier heat law is one of the most
important properties in the nonequilibrium thermodynamics. 
It states that the heat current per volume is proportional to the
thermal gradient. 
Microscopic dynamical origin for the realization of Fourier heat law
has been actively studied using many Hamiltonian systems 
\cite{RLL67,LLP97,STM96}.
In the complete harmonic chain, no global thermal gradient appears,
and local equilibrium is not realized, which are attributed to the lack of
scattering between modes \cite{RLL67,STM00}.
In an isotropic $d$-dimensional classical 
Fermi-Pasta-Ulam (FPU) system which has the nonlinear potential term, 
the mixing property satisfies 
because the auto-correlation function of the energy current roughly 
decays as $t^{-d/2}$. However due to its slow relaxation
of the current fluctuation, the Green-Kubo formula diverges in the one- and 
two-dimensional systems which causes an anomalous energy 
transport \cite{LLP97,LL00}. 

In low dimensional systems, most of the problems preventing a normal
thermal conduction arises from a slow fluctuation of energy current,
including the failure of mixing property. 
This situation is also the case in the quantum systems.
Many integrable one-dimensional systems show the failure of mixing property.
In the isotropic Heisenberg chain, the energy current operator commutes
with the Hamiltonian, which trivially causes the failure of mixing
property and the abnormal energy transport \cite{MZ}.
The recent experiments confirmed a such ballistic heat 
transport in Sr$_2$CuO$_{3}$ \cite{SFGOVR00}
and CuGeO$_{3}$ \cite{cugeo3} which are described by the 
isotropic Heisenberg chain.

In this paper, we study how the normal thermal conduction in a one-dimensional 
quantum system is realized in microscopic point of view.
In quantum systems, the dynamical origins 
of normal thermal conduction and the related quantum effects are not 
enough understood\cite{STM96,P98}. In classical systems, {\em chaos}
characterized by the sensitivity to the initial state can be induced 
by nonlinearities and can play a crucial role for the mixing
property. However such {\em chaos} cannot be induced by a 
linear Schr\"{o}dinger equation in quantum systems \cite{JP96}. 
In spite of such discrepancies in dynamics, 
the Fourier heat law can be realized in quantum systems.
We show the strong evidence of the normal 
thermal conduction in a one-dimensional quantum spin system
by direct investigation of properties under thermal gradient and 
the Green-Kubo formula.
We find that the autocorrelation function 
in the Green-Kubo formula approximately decays as $t^{-1.5}$.
Thus the mixing property is satisfied and the formula converges.

We consider the tractable simple quantum system whose Hamiltonian is described as,
\begin{eqnarray}
{\cal H} &=& \sum_{\ell =1}^{N_1} J \sigma_{\ell}^{z}\sigma_{\ell +1 }^{z}
+D \left( \sigma_{\ell}^{z} \sigma_{\ell +1 }^{x}
-\sigma_{\ell}^{x} \sigma_{\ell +1 }^{z} \right)   
+  \Gamma\sum_{\ell =1}^{N}\sigma_{\ell}^{x}
+ H\sum_{\ell=1}^{N} \sigma_{\ell}^{z} , \label{hamil}
\end{eqnarray}
where $\sigma_{\ell}^{\alpha}$ $(\alpha = x,y,z)$ 
is the $\alpha$ component of the Pauli
matrix at the $\ell$th site. 
$N$ is the number of spin, and $N_1$ is taken as $N-1$ for the open
boundary condition and $N$ for the periodic boundary condition, respectively.
The first term is the nearest neighbor Ising interaction, and
the second term is the Dzyaloshinsky-Moriya (DM) interaction, where
the anisotropic vector is taken as ${\bf D} = (0, D, 0)$ for  
the general form 
${\bf D}\cdot \left( {\bf\sigma}_{\ell}\times {\bf\sigma}_{\ell +1}
\right)$.
The third and fourth term are the Zeeman term of $x$ and $z$ direction, 
respectively. 

This Hamiltonian shows the variety of symmetries 
by controlling parameters $(J,D,\Gamma, H)$.
In the case of periodic boundary condition, there exists the 
translational symmetry $\sigma_{n} \leftrightarrow \sigma_{n+\ell} 
\, (\ell = 1, \cdots ,  N)$ .
In the absence of the DM interaction, i.e, $D=0$, the system has 
the reflection symmetry, $\sigma_{n} \leftrightarrow \sigma_{N-n}$. 
In the case of $D=0$ and $H=0$,
the system Hamiltonian is diagonalized by using free fermions 
through the Jordan-Wigner transformation \cite{K62}. 
When $\Gamma=0$ and $H=0$, the system 
has the time reversal symmetry which yields the Kramers degeneracies.
In many systems with symmetries, conserved quantities
prevent a normal heat transport, so that the Green-Kubo formula
diverges \cite{MZ,note2}. 

As well known, level spectrum shows the universal statistics
according to the nonintegrability of system\cite{BGS84}. 
We investigate the cumulative spacing distribution $I(S)$
defined by $I(S) = \int_{0}^{S} \, d u \, P(u) $
for the energy spacing distribution $P(u)$.
We show the distributions in Fig.1 for $(J,D,\Gamma,H)=(0.2,0.2,0.2,-0.2)$ 
and $(0.2,0,0.2,0)$ with the open boundary condition $N_1 = N-1$. 
The former system has no trivial symmetries, while the latter can be
mapped into free-fermion. Points are numerical
data by diagonalizing the Hamiltonian (\ref{hamil}), and the lines 
are theoretical predictions, i.e., 
the Poisson distribution $I(S) = 1- \exp(-S)$ for integrable systems, 
and the Wigner distribution $I(S) = 1-\exp(-\pi S^2 /4)$ 
for nonintegrable systems, respectively. The figure for 
the parameters $(0.2,0.2,0.2,-0.2)$ shows the agreement between 
the numerical data and the Wigner surmise. 
For small spacing $S\ll 1$, $I(S)$ behaves as $I(S) \propto S^{1.92\pm 0.02}$, 
which is very close to the Wigner distribution $I(S) \propto S^{2}$. 
This universal feature of spectrum indicates a complex dynamics of this system.
Thus throughout this paper, we take the parameters as
$J=D=\Gamma = -H = 0.2$ \cite{note}.
\begin{figure}[t]
\begin{center}
\noindent
\epsfxsize=8.7cm \epsfysize=5.8cm
\epsfbox{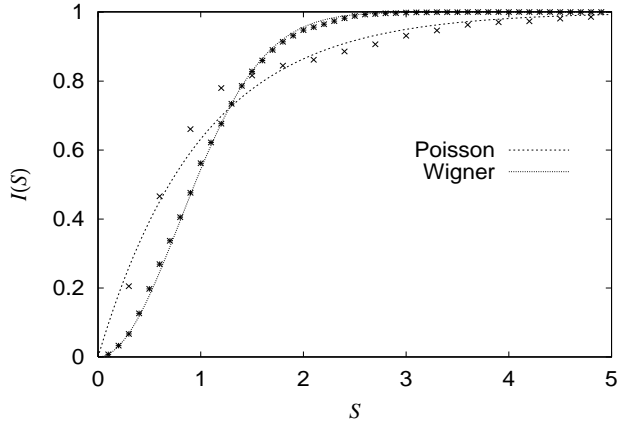}
\caption{Cumulative level spacing distribution for $N=12$ with the open
 boundary condition. The asterisk and cross are for the cases of 
$(J,D,\Gamma,H)=(0.2,0.2,0.2,-0.2)$ and $(0.2,0,0.2,0)$, respectively.
For small $S$ in the former case, $I(S)$ is approximately fitted by 
$I(S) \propto S^{1.92\pm 0.02}$.}
\end{center}
\end{figure}

In order to investigate properties of the energy transport 
under thermal gradient, the reservoirs of different 
temperatures are directly connected with the system at the both ends.
Here we adopt the phonon reservoir with the Ohmic spectral density 
for which the master equation of the system is written as (see
e.g., \cite{STM00} and references therein);
\begin{equation}
\frac{\partial\rho(t)}{\partial t} = 
-i\left[ {\cal H}, \rho (t) \right] -\lambda 
\left( {\cal L}_{\rm L} \rho (t)+ {\cal L}_{\rm R} \rho (t) \right) ,
\label{QMEq}
\end{equation}
where the first term in the right-hand side corresponds to 
the pure quantum dynamics of the system, and 
${\cal L}_{\rm  L}$ and ${\cal L}_{\rm  R}$ express
the dissipative effects of the inverse temperature $\beta_{\rm L}$ at
the left reservoir and  $\beta_{\rm R}$ at the right reservoir,
respectively. The parameter $\lambda$ is the coupling strength.
The dissipative term $ {\cal L}_{\alpha } \, (\alpha =
{\rm L},{\rm R})$ is given by
$
 {\cal L}_{\alpha } \rho(t) =
\left( \left[X_{\alpha }, R_{\alpha } \rho(t)\right] 
+ \left[X_{\alpha }, R_{\alpha } \rho(t) \right]^{\dag} \right) ,
$
where $X_{\rm L}$ and $X_{\rm R}$ are the system's operators 
directly attached to the left and right reservoir, respectively.
Here we take $X_{\rm L} = \sigma_{1}^{z}$ and $X_{\rm R } =
\sigma_{N}^{z}$.
The operator $R_{\alpha }$ is given by 
$\langle k | R_{\alpha }  | m \rangle = 
(E_{k} - E_{m}) ({e^{\beta_{\alpha } (E_{k} - E_{m})} -1 })^{-1}
\langle k | X_{\alpha }  | m \rangle $ in the representation 
diagonalizing the Hamiltonian (\ref{hamil}) as ${\cal H} | k \rangle =
E_{k} | k \rangle $ and ${\cal H} | m \rangle = E_{m} | m \rangle $.

We first consider the energy profile defining the $\ell$th local 
Hamiltonian ${\cal H}_{\ell}$;
\begin{eqnarray}
{\cal H}_{\ell} = J \sigma_{\ell}^{z}\sigma_{\ell +1 }^{z} +
D \left( \sigma_{\ell}^{z} \sigma_{\ell +1 }^{x} 
-\sigma_{\ell}^{x} \sigma_{\ell +1 }^{z} \right)
+ \sum_{k = \ell}^{\ell +1} \left(
\Gamma \sigma_{k}^{x} + H \sigma_{k}^{z} \right) \label{lsb}.
\end{eqnarray}
We numerically integrate the equation (\ref{QMEq})  
and obtain the stationary density matrix $\rho_{\rm st}$ when 
the initial density matrix is taken as the canonical distribution of 
temperature $1/\beta_{\rm R}$.
The simulation was carried out 
by the fourth order Runge-Kutta method with the time step $0.01$ 
for the parameters $\beta_{\rm L} =  0.5$, $\beta_{\rm R} =  0.2$, 
and $\lambda =0.01$.
In Fig. 2, we show the energy profiles at the stationary
state ${\rm Tr}\left(\rho_{\rm st}{\cal H}_{\ell} \right)$ 
for $N=7,8,9$, and $10$. 
The two large circles are equilibrium energy 
values of inverse temperatures $\beta_{\rm L}$ and 
$\beta_{\rm R}$, respectively. 
$N=10$ was the largest number within the present computational ability.
The figure shows the profiles with a finite gradient
which is the {\em sufficient condition} for the Fourier heat law,
although we cannot enumerate the energy profile and energy current 
in the thermodynamic limit.
The local energies at the both edges are different
from the expectation values of the temperatures of reservoirs. 
This is attributed to the small coupling strength 
$\lambda (=0.01)$. Actually we confirmed that 
for larger $\lambda \,\, (\, \ll J,\, D)$, the energy profile 
smoothly changes and these differences become smaller. 
We found the qualitatively same finite energy gradient in 
the case of smaller temperature difference $\beta_{\rm L } = 0.3$ 
and $\beta_{\rm R} = 0.2$.
\begin{figure}[t]
\begin{center}
\noindent
\epsfxsize=8.7cm \epsfysize=5.8cm 
\epsfbox{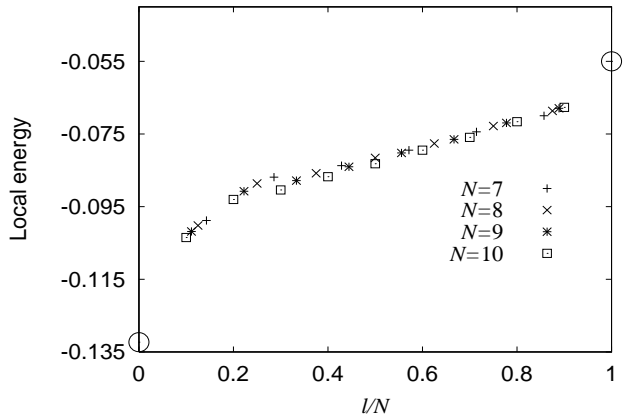} \\
\caption{The energy profile at the stationary state for from $N=7$
 to $N=10$. The abscissa is the scaled spin number, $l /N , l = 1, 
\cdots ,N-1$. }
\end{center}
\end{figure}

We next study the local energy distribution at the stationary state
focusing on whether the local equilibrium is realized or not for 
the system of $N=10$. 
The local equilibrium can be realized even in finite size 
systems\cite{STM99,D01}. 
We study the energy distribution of subsystems 
composed of two spins and four spins.
We consider four local subsystems with two spins, i.e., 
the subsystems described by the local Hamiltonians,  
${\cal H}_{2}, {\cal H}_{4}, {\cal H}_{6}$, and $ {\cal H}_{8} $,
and two local subsystems with four spins, i.e., 
the subsystems described by 
$({\cal H}_{2} + {\cal H}_{3}) $ and $({\cal H}_{6} + {\cal H}_{7})$.
Temperatures at boundaries are set as 
$(\beta_{\rm L}, \beta_{\rm R})=(0.5,0.2) $ for the former case
and $(\beta_{\rm L}, \beta_{\rm R})=(0.3,0.2)$ for the latter case.
The $\ell$th local energy distribution $P_{\ell}(\varepsilon_i) $ is 
calculated using the $\ell$th local reduced density matrix $\rho_{\ell} $ as,
\begin{eqnarray}
P_{\ell} (\varepsilon_i) &=& 
\langle i_{\ell} | \rho_{\ell} |i_{\ell} \rangle ,
\end{eqnarray}
where $|i_{\ell}\rangle$ is the $i$th eigenstate of the $\ell$th 
local subsystem Hamiltonian. 
The numbers of eigenstates are $4$ and $16$ for 
 the cases of $(\beta_{\rm L}, \beta_{\rm R})=(0.5,0.2) $ 
and $(\beta_{\rm L}, \beta_{\rm R})=(0.3,0.2)$, respectively.
The $\ell$th local density matrix $\rho_{\ell} $ is obtained by 
taking the trace for stationary density matrix $\rho_{\rm st}$ 
in the Hilbert space exclusive of spins which belong to the $\ell$th subsystem.
In Fig.3, the energy distributions $P_{\ell} (\varepsilon_i)$ are shown
on the semi-log scale.
The points linked by one line are the distribution at one local subsystem.
With the increase of the subsystem number $\ell$, the gradient of the line 
becomes smaller. The distributions approximately take the exponential form. 
Thus this monotonic $\ell$ dependence of gradient of energy distribution
indicates that this system can be the {\em candidate} which satisfies 
the local equilibrium. 
We must note that the Ising and DM interaction 
between nearest neighbor local subsystems and the existence 
of finite energy flow can cause deviations of $P_{\ell} (\varepsilon_i)$ 
from the exponential form. 
\begin{figure}[t]
\begin{center}
\noindent
\epsfxsize=13.3cm \epsfysize=5.5cm
\epsfbox{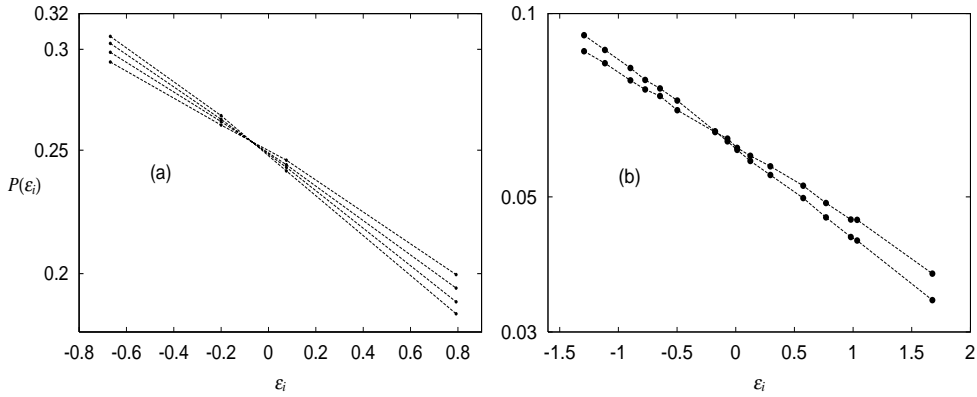}
\caption{ Local energy distribution for size $N=10$.
(a) Four local subsystems composed of two spins for $(\beta_{\rm L}, 
\beta_{\rm R})=(0.5,0.2) $. 
(b) Two local subsystems composed of four spins
for $(\beta_{\rm L}, \beta_{\rm R})=(0.3,0.2) $. }
\end{center}
\end{figure}

Generally even if the energy profile is well scaled and the local
equilibrium is realized, thermal conductivity 
can diverge in the thermodynamic limit due to the slow relaxation of
energy currents mentioned in the introduction \cite{D01}. 
We now investigate whether this type of divergence occurs or not 
in the present system. The Green-Kubo formula is 
derived on the basis of the local equilibrium state. 
Since the present system is the candidate which satisfies 
the local equilibrium,
we expect that the Green-Kubo formula quantitatively 
describes the thermal conductivity $\kappa (\beta )$ 
in the thermodynamic limit. The Green-Kubo formula reads as,
\begin{eqnarray}
\kappa (\beta ) &=& 
\frac{\beta^2}{2N} \int_{0}^{\infty} \, du \, 
\langle\left\{ \hat{J}, \hat{J}(u)\right\}\rangle ,
\label{gkf}
\end{eqnarray}
where $\left\{ ., .\right\}$ means the anti-commutation relation, and 
$\langle ... \rangle$ is the equilibrium average at the inverse
temperature $\beta$.
The operator $\hat{J}(u)$ is the total current operator at the time $u$ 
in the Heisenberg picture, i.e., 
$\hat{J}(u) = \exp(i {\cal H} u) \hat{J} \exp(-i {\cal H} u)$. 
The total current operator $\hat{J}$ is calculated
by the continuity equation of energy as
$
\hat{J} = -i \sum_{\ell}\left( 
\left[ \hat{h}_{\ell +1}, \hat{h}_{\ell} \right]  +
\left[ \hat{h}_{\ell +2}, \hat{h}_{\ell} \right]
\right) , $ 
using 
$ 
\hat{h}_{\ell} = J \sigma_{\ell}^{z}\sigma_{\ell +1 }^{z} +
D \left( \sigma_{\ell}^{z} \sigma_{\ell +1 }^{x} 
-\sigma_{\ell}^{x} \sigma_{\ell +1 }^{z} \right)
+ \Gamma \sigma_{\ell}^{x} + H \sigma_{\ell}^{z}. $
In order to consider the thermodynamic limit, 
we define $A(t)$ and $C(t)$ as follows,
\begin{eqnarray}
\begin{array}{ccc}
A(t)=  \frac{1}{2N}  \langle \left\{ \hat{J}, \hat{J}(t) \right\} \rangle, 
&\mbox{}&
C(t) = \int_{0}^{t} \, du \, A(u) ,
\end{array}
\end{eqnarray}
where $A(t)$ is the autocorrelation function for the total current, and 
$C(t)$ is the integration of $A(t)$. We call $C(t)$ the Green-Kubo integral.
The numerical calculations are carried out for $\beta = 0.3$ 
with the periodic boundary condition $N_1 = N$. 
In Fig. 4(a), we plotted the absolute values of the autocorrelation 
functions $A(t)$ for system sizes $N=12, 14$ and $16$ on the log-log scale. 
Up to $N=14$, we used the numerical diagonalization for the Hamiltonian, 
and for $N=16$ we calculated the time evolution of wave function for 
the randomly chosen $1024$ initial states \cite{P98}. 
The figure indicates the power law decay $\sim t^{-1.5}$ of $A(t)$. 
In the inset, the raw data of $A(t)$ for various system sizes 
$(N=8,10,12,14$, and $16)$ are plotted on the normal scale.
All data are saturated to some finite 
values with fluctuations. This constant is partially caused by 
the energy degeneracies due to translational symmetry of periodic condition.
Thus the behavior of $A(t)$ is roughly represented as follows,
\begin{eqnarray}
A(t) \sim a(N,t) \, t^{-1.5} + b(N,t)+ B (N),
\end{eqnarray}
where $a(N,t)$ and $b(N,t)$ are some fluctuating functions 
with a vanishing mean value and  
$B(N)$ denotes the saturated average value for $N$. 
$B(N)$ can be exactly calculated by a numerical diagonalization up to $N=14$.
For $N=16$, it is calculated by directly averaging the numerical data of
$A(t)$. 
We find that $B(N)$ vanishes with the increase of $N$ 
following or faster than the exponential function 
$0.1\times \exp \left( -0.5 N \right)$,
as shown in Fig. 4(b). There we plotted the data from $N=6$ to $N=16$.
Thus the integral $C(t)$ is expected to converge showing 
roughly the following behavior in the thermodynamic limit $N\rightarrow \infty$,
\begin{eqnarray}
C(t) \sim \int^{t} \, a(N,u) u^{-1.5} \, du + B(N) t
\rightarrow  \int^{t} \, a(\infty ,u) u^{-1.5} \, du < C_0 + t^{-0.5} .
\end{eqnarray} 
In Fig.5, we present the numerically calculated $C(t)$ for various $N$. 
The behavior of $C(t)$ strongly supports the convergence of the
Green-Kubo integral in the thermodynamic limit. 
The behavior of $C(t)$ reminds us the order of limit, i.e., 
taking the limit $N$ to infinity first and then $t$ to infinity. 
From the convergence of $C(t)$, we conclude that the present one-dimensional 
spin system without trivial symmetries shows the finite 
thermal conductivity which is independent of the system 
size as far as $N$ is very large.

\begin{figure}[t]
\begin{center}
\noindent
\epsfxsize=7.0cm \epsfysize=4.8cm
\epsfbox{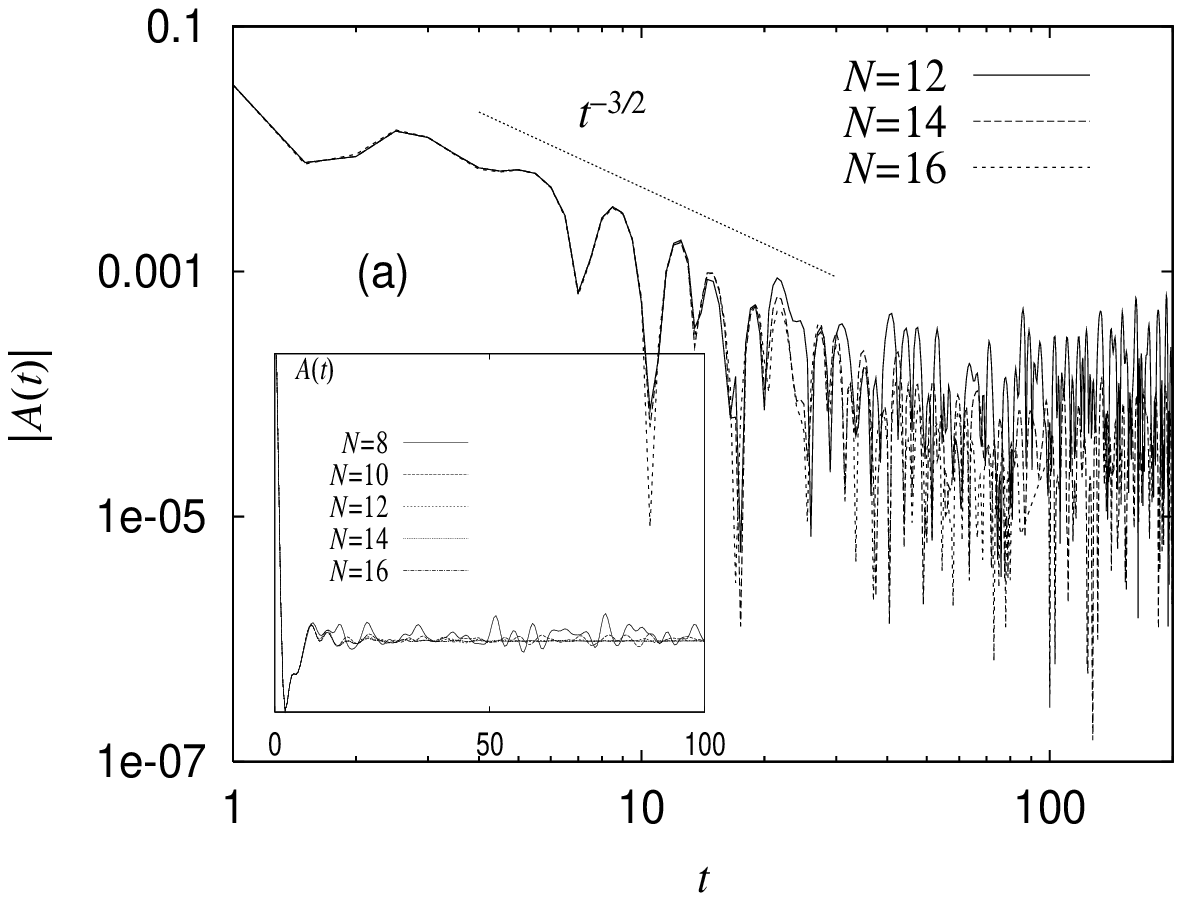} 
\epsfxsize=7.0cm \epsfysize=4.8cm
\epsfbox{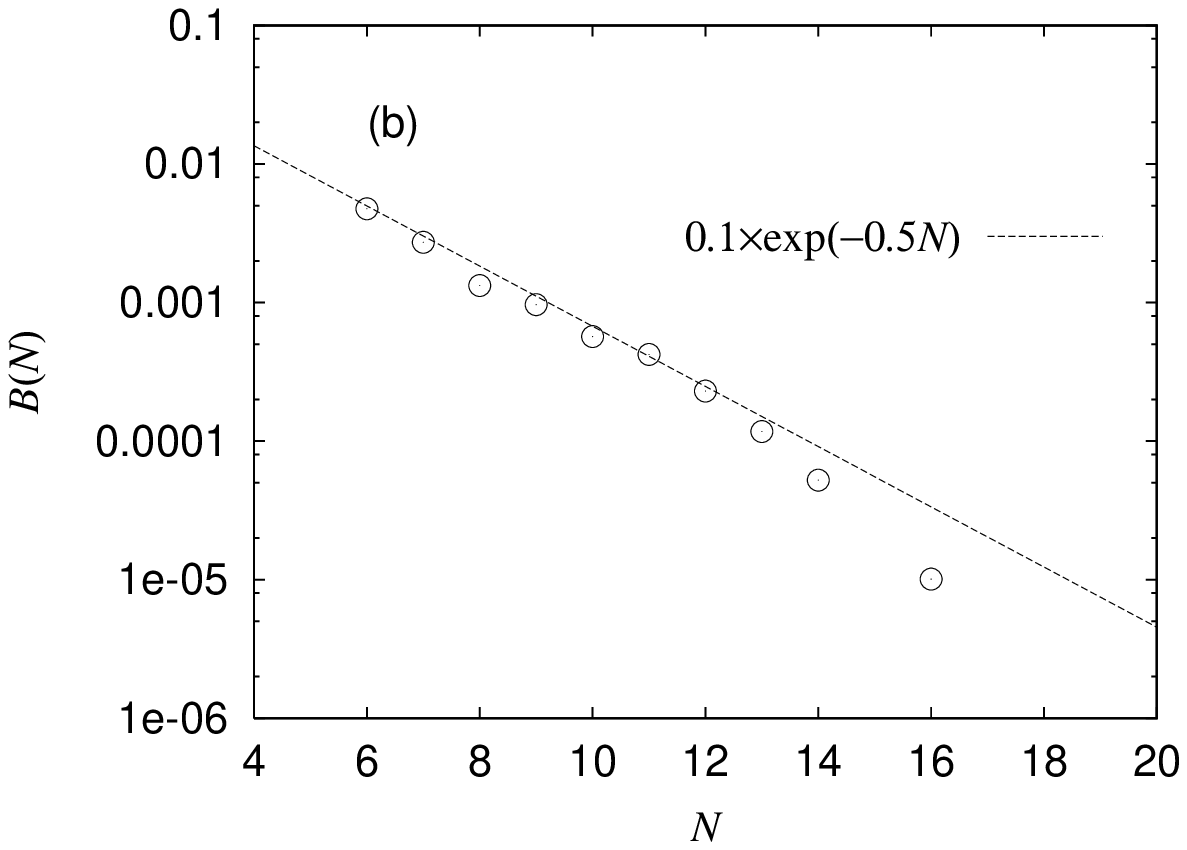}
\caption{(a): The absolute values of autocorrelation function as
 function of time for $N=12,14$, and $16$. In the inset,
the raw data are shown for $N=8,10,12,14$, and$16$. 
(b): The average constant $B(N)$ as a function  of $N$. 
The dashed line is the exponential function 
$0.1\times \exp(-0.5\times N)$. }
\end{center}
\end{figure}

\begin{figure}[t]
\begin{center}
\noindent
\epsfxsize=8.5cm \epsfysize=5.8cm
\epsfbox{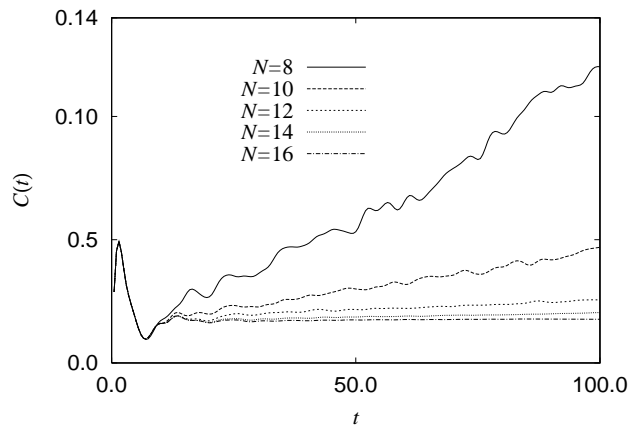} \\
\caption{The Green-Kubo integral $C(t)$ as a function of time 
for $N=8,10.12,14$, and  $16$.}
\end{center}
\end{figure}

In conclusion, we find the strong evidence of normal thermal conduction 
in the present system by microscopic investigation.
The autocorrelation function in the Green-Kubo formula shows 
power law decay $\sim t^{-1.5}$, which may be the characteristics 
in the high temperature region, i.e. semiclassical region \cite{P00}. 
It is also interesting to study the behavior at very low temperatures. 
The Fig.3 indicates that this system can be 
the candidate which satisfies 
the local equilibrium. We should carefully and systematically investigate
how the exponential form in the local 
energy distribution are realized. 

In the recent experiments showing a ballistic heat 
transport in of Sr$_2$CuO$_{3}$ \cite{SFGOVR00}
and CuGeO$_{3}$ \cite{cugeo3},
the energy is transmitted by spin excitations (magnon), and 
the coherent length is very long (about $100$ times lattice constant
in CuGeO$_{3}$ above the spin-Peierls temperature).
Theoretical studies for temperature dependence of thermal conductivity 
are actively done \cite{KS02,SM02}.
It is also interesting to investigate theoretically and
experimentally how the coherent length changes 
when the symmetries disappears in these realistic systems. 


\acknowledgments

The author would like to thank S. Takesue and T. Shimada for 
valuable discussion. He also would like to thank the valuable comments
of the referees. The computer calculation 
was partially carried out at the computer
center of the ISSP, which is gratefully acknowledged.
The present work is supported by Grand-in-Aid for Scientific 
Research from Ministry of Education, Culture, Sports, Science,
and Technology of Japan.



\begin{thebibliography}{0}
\bibitem{note2}
For example, in the free fermion case of $D=H=0$, the fermion can run
	through without scatterings, which causes the ballistic transport.
In that case,  the temperature profile becomes flat \cite{STM96}.
\bibitem{note} 
We regard the system (\ref{hamil}) as a dynamical model which shows 
a variety of complex dynamics. 
We take this simple set of parameters although the DM interaction
is relatively large.
\end{thebibliography}
\end{document}